\documentclass[12pt,preprint]{aastex}

\usepackage{amsmath}
\usepackage{graphicx}
\usepackage{amssymb}

\newcommand{\kms}{km\,s$^{-1}$}

\newcommand{\myemail}{liuchao@nao.cas.cn}

\slugcomment{}

\shorttitle{A resonant feature near the Perseus arm}
\shortauthors{Liu et al.}

\begin{document}


\title{A resonant feature near the Perseus arm revealed by red clump stars}


\author{Chao Liu\altaffilmark{1,2}, Xiangxiang Xue\altaffilmark{1,2}, Min Fang\altaffilmark{3,4}, Glenn van de Ven\altaffilmark{2}, Yue Wu\altaffilmark{1},  Martin C. Smith\altaffilmark{5,6}, Kenneth Carrell\altaffilmark{1}}
\email{\myemail}
\altaffiltext{1}{Key Lab of Optical Astronomy, National Astronomical Observatories, CAS, 20A Datun Road, Chaoyang District, 100012 Beijing, China}
\altaffiltext{2}{Max Planck Institute for Astronomy, K\"onigstuhl 17, D-69117 Heidelberg, Germany}
\altaffiltext{3}{Pulple Mountain Observatory, 2 West Beijing Road, 210008, Nanjing, PR China}
\altaffiltext{4}{Key Laboratory of Radio Astronomy, Chinese Academy of Sciences}
\altaffiltext{5}{National Astronomical Observatories, Chinese Academy of Sciences, 20A Datun Road, Chaoyang District, 100012 Beijing, China}
\altaffiltext{6}{Kavli Institute for Astronomy and Astrophysics, Peking University, 100871 Beijing, China }

\begin{abstract}
We investigate the extinction together with the radial velocity dispersion and distribution of red clump stars in the anti-center direction using spectra obtained with Hectospec on the MMT. We find that extinction peaks at Galactocentric radii of about 9.5 and 12.5\,kpc, right in front of the locations of the Perseus and Outer arms and in line with the relative position of dust and stars in external spiral galaxies. The radial velocity dispersion peaks around 10\,kpc, which coincides with the location of the Perseus arm, yields an estimated arm-interarm density contrast of 1.3--1.5 and is in agreement with previous studies. Finally, we discover that the radial velocity distribution bifurcates around 10--11\,kpc into two peaks at $+27$\,\kms\ and $-4$\,\kms. This seems to be naturally explained by the presence of the outer Lindblad resonance of the Galactic bar, but further observations will be needed to understand if the corotation resonance of the spirals arms also plays a role. 
\end{abstract}


\section{Introduction}
\label{s:intro}

Non-axisymmetric rotating substructures, such as a bar and spiral arms in the Milky Way, induce a corotation resonance (CR) as well as inner (ILR) and outer Lindblad resonances (OLR) around certain Galactocentric radii \citep[e.g.,][]{weinberg94,fux00,quillen11,antoja11}.
These resonances seem to be able to explain the moving groups in the velocity distribution in the Solar neighborhood discovered from Hipparcos data and the Geneva-Copenhagen Survey \citep[e.g.,][]{dehnen00,desimone04,quillen05,sellwood09}. 
However, resonances from multiple patterns may exist and overlap, and moreover the Sun may not be located right on any of the resonance radii. Henceforth, it is difficult to distinguish the effects of resonances in the Solar neighborhood. Observations of the velocity distribution outside the Solar radius are needed to constrain the resonances.

We report in this letter a resonance feature at a Galactocentric radius of 10--11\,kpc in the radial stellar velocity distribution derived from MMT/Hectospec observations of $698$ red clump stars in the anti-center direction. In Section~\ref{s:data} we briefly describe the observations, data reduction and analysis. In Section~\ref{s:discussion} we identify the location of the Perseus arm and discuss possible reasons for the bifurcation seen in the radial velocity distribution. Finally, we summarize our findings in Section~\ref{s:summary}.

\section{Data reduction and analysis}
\label{s:data}

\begin{figure}[t!]
\begin{center}
\includegraphics[width=14cm]{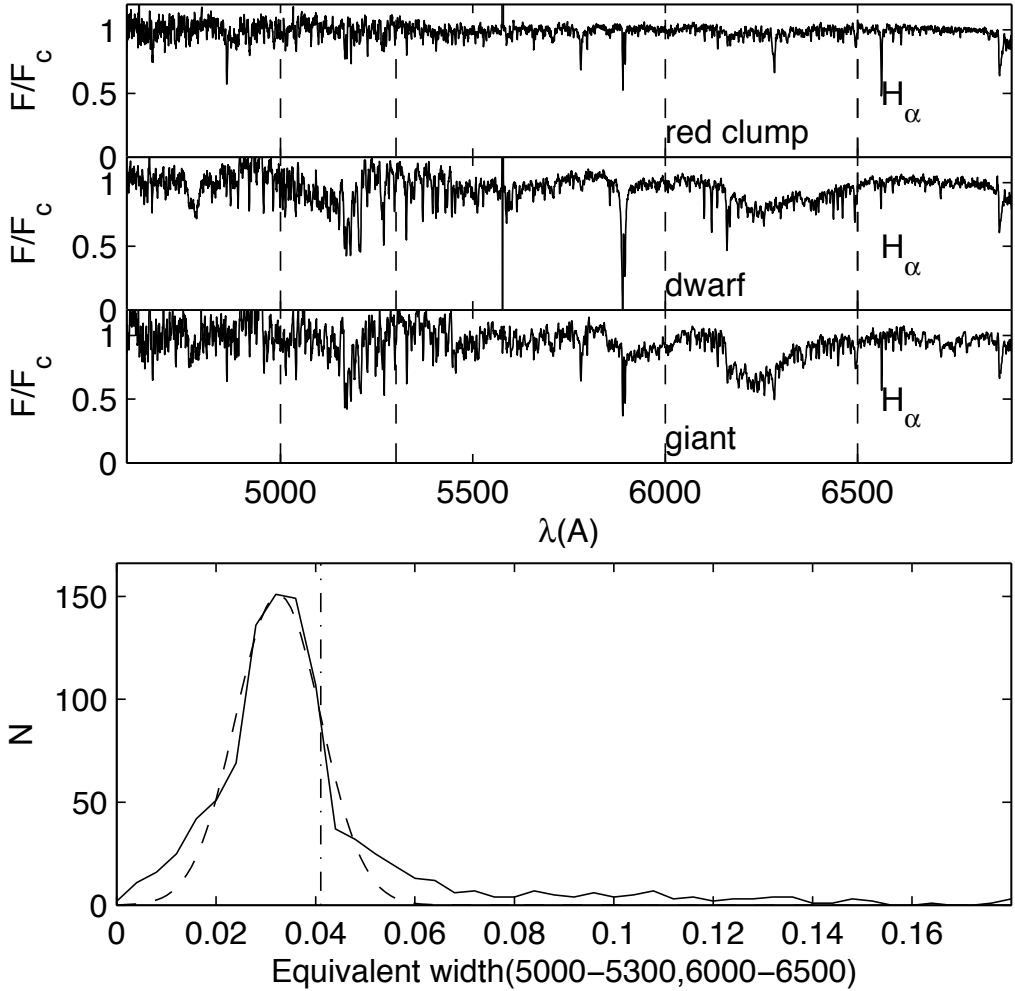}
\caption{\emph{Top panel: }Reduced MMT/Hectospec spectra of a red clump star, dwarf star, and giant star. The equivalent width between the pairs of vertical dashed lines at $5000\rm\AA$--$5300\rm\AA$ and $6000\rm\AA$--$6500\rm\AA$ is used to remove the dwarf and giant star contaminants based on their prominent MgH and TiO bands. \emph{Bottom panel: } The solid curve shows the distribution of the EW for all spectra. The dashed curve is the Gaussian fit to the peak with the dot-dashed vertical line indicating one standard deviation above the mean, above which stars are excluded from the red clump sample to safely remove all the dwarf and giant star contaminants in the tail.}
\label{fig:sample}
\end{center}
\end{figure}

We used MMT/Hectospec to obtain spectra with resolution R=2500 of stars in 4 one-degree-diameter fields between $l=178^\circ$--$182^\circ$ at $b=0^\circ$. Red clump stars are selected as targets from the 2MASS catalog by a single color cut $J-K>0.65$\,mag. Close-by dwarf stars can overlap in intrinsic color with the red clump stars, but as the latter get reddened by extinction in the mid-plane the dwarf stars can be easily removed. The single-color selection of red clump stars is thus expected to be very efficient with only little contamination by other giant stars and remaining dwarf stars. 

In total we obtained 993 spectra with mean $S/N\simeq17$. The reduced spectra, of which three examples are shown in the top panel of Fig.~\ref{fig:sample}, show that later-type K/M giant stars and dwarf stars contain stronger MgH and TiO absorption features within $5000\rm\AA$ --$5300\rm\AA$ and $6000\rm\AA$--$6500\rm\AA$ than red clump stars. After computing the equivalent width (EW) of the absorption features in each spectrum, we arrive at the distribution in the bottom panel of Fig.~\ref{fig:sample} with a Gaussian peak of most, if not all, red clump stars plus a tail of higher EW caused by the MgH and TiO bands in the contaminating giant and dwarf stars. After conservatively selecting the stars with EW not larger than one standard deviation above the mean, we are left with 747 red clump stars. This implies a completeness of at least 85\% and hence confirming the high effectiveness of the color selection. 
To enable the derivation of a smooth radial distribution in extinction and velocity below, we exclude another 46 stars with excessive local extinction as well as 3 stars that are 3-$\sigma$ outliers in velocity, leaving a final sample of 698 red clump stars. 
Three independent methods, $H_\alpha$ line fitting, synthetic template matching and ULySS \citep{wu11}, are used to extract a mean radial velocity value and corresponding error from each spectrum. 

Next, we use the Hipparcos \citep{perryman97} and 2MASS \citep{skrutskie06} catalog to infer, after taking into account photometric and parallax errors, the intrinsic $J-K$ color index and absolute $K$-band magnitude of the red clump stars: $(J-K)_0=0.654\pm0.058$\,mag and $M_K=-1.547\pm0.311$\,mag. The latter is in agreement with both \citet{groenewegen08} and \citet{lopez02}, and since metallicity is at most weakly correlated with $M_I$ \citep{zhao01}, and likely less so for $M_K$, we can obtain accurate distance estimates to the red clump stars in the following way.

For each red clump star the probability density function (PDF) of extinction given its observed color index, $p(A_K|J-K)$, is computed, assuming it is a Gaussian distribution with dispersion given by the previously derived intrinsic $J-K$ dispersion plus the $J-K$ photometric error added in quadrature. 
Similarly, the likelihood of apparent $K$-band magnitude $K$ given the distance modulus $DM$ and $A_K$, $p(K|DM, A_K)$, is assumed to follow a Gaussian distribution with dispersion given by the previously derived intrinsic $M_K$ dispersion together with the photometric uncertainty in $K$.
The PDF of the distance modulus given the observed $J-K$ color and $K$ then follows from Bayesian's theorem as
\begin{multline}
\label{eq:DMestiamte}
	p(DM|K, J-K) = \int p(K|DM,A_K)p(A_K|J-K) \, p(A_K) \, p(DM) \, \textrm{d} A_K,
\end{multline}
with adopted uniform priors $P(A_K)$ and $P(DM)$. Adopting as Solar radius $R_\odot=8$\,kpc, the distance modulus DM is converted to Galactocentric radius $R_{GC}$.

\begin{figure}[t!]
 \vspace{-10pt}
\begin{center}
\includegraphics[width=7cm]{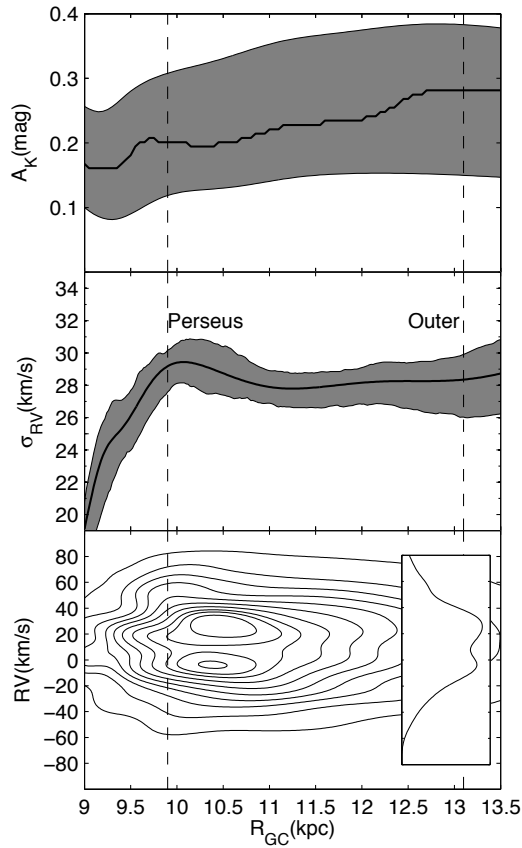}
\caption{
Structural and velocity distribution of red clump stars as a function of Galactocentric radius $R_{GC}$ with the location of the Perseus and Outer arms indicated. 
\emph{Top panel: } Most likely $K$-band extinction $A_K$ and its 68\% confidence region.
\emph{Middle panel: } Same for the radial velocity dispersion $\sigma_{RV}$.
\emph{Bottom panel: } The probability of radial velocity versus distance with 95\%, 90\%, \dots, 5\% contours from outside to inside. The inset shows the distribution marginalized over the 10--11\,kpc range in $R_{GC}$.
}
\label{fig:AkUsigU_D_norm}
\vspace{-5pt}
\end{center}
\end{figure}

Fig.~\ref{fig:AkUsigU_D_norm} shows the distributions of K-band extinction $A_K$, radial velocity dispersions $\sigma_{RV}$ and radial velocity ($RV$) versus $R_{GC}$ distribution. Given a distance, the distribution in $A_K$ follows by summing up  $p(A_K|J-K)p(DM|K,J-K)$ for all stars. The line in the top panel in Fig.~\ref{fig:AkUsigU_D_norm} shows the $A_K$ with maximum probability. And the shadow shows the 68\% confidence region. 
%
There are two significant jumps and hence dense dust regions around 9.5\,kpc and 12.5\,kpc is found in the distribution in $A_K$. 

The radial profile of $\sigma_{RV}$ in the middle panel is obtained from the weighted variance of RV, which equals to $\sigma_{RV}^2$ plus the squared measurement error of RV, $\epsilon_{RV}^2$. The weight is given by $p(DM|K,J-K)$ for each star with normalization. During the calculation we assume that the probability of the measured RV is a Gaussian with sigma $\epsilon_{RV}^2$. The uncertainty of $\sigma_{RV}$ is given from a bootstrap resampling process.  It reveals that there is a clear bump at 10\,kpc.

Finally, the bottom panel shows the radial velocity ($RV$) versus $R_{GC}$ distribution from combining the PDFs in radial velocity and distance for each star. The uncertainties in $RV$ are about 3.2\,km s$^{-1}$ and in $R_{GC}$ are about 14.6\% and smooth the total distribution. Even so, the velocity distribution shows two significant peaks around 10--11\,kpc at $RV$s of about $+27$\,\kms\ and $-4$\,\kms.

\section{Discussion}
\label{s:discussion}

\subsection{Dusty arms}
\label{ss:dust}

The jump in extinction around  9.5\,kpc in the top panel of Fig.~\ref{fig:AkUsigU_D_norm} indicates a broad and dense dust region in the path of the line-of-sight. Its location is just short of the Perseus arm at 9.9\,kpc \citep{reid09} implying that it is the dustlane of the spiral structure. Indeed, taking into account the Galactic rotation direction, the dustlane is in front of the stellar arm. Similarly, the jump in extinction around 12.5\,kpc is in front of the Outer arm located at 13.1\,kpc \citep{reid09}. 

\subsection{Arm-interarm contrast}
\label{ss:contrast}

In the anti-center direction the measured radial velocity $RV$ is approximately equal to the Galactic space or cartesian velocity $U$, but with opposite sign. The bump in $\sigma_{RV}$ around 10\,kpc indicates the mass center of the Perseus arm. Since the velocity dispersion in a equilibrium system is related to the mass density, the arm--interarm density contrast can be estimated. 
The peak value of $\sigma_{RV}$ is 29.4\,\kms\ at 10\,kpc, while it decreases to 27.8\,\kms\ at the interarm radius of around 11\,kpc. Assuming an isothermal density distribution for the stars and a constant vertical-to-radial velocity dispersion ratio $\sigma_z/\sigma_R$, $\sigma_R$ is proportional to $\Sigma^{1/2}$ where $\Sigma$ is the surface density. The arm--interarm contrast is then $K\simeq1.3$\footnote{Here $K=(\Sigma+\delta\Sigma)/(\Sigma-\delta\Sigma)$, in which $\delta\Sigma$ is the incremental surface density in the arm, following \citet{antoja11}}. Alternatively, if $\sigma_R$ is proportional to $\Sigma^{1/3}$ \citep{dehnen99}, we obtain $K\simeq1.5$. The resulting contrast of $K\simeq1.3-1.5$ is in agreement with previous studies \citep{drimmel01,benjamin05,grosbol04}.


%

\subsection{Bifurcation due to a resonance?}
\label{ss:bifurcation}

The bifurcation in $RV$ around 10--11\,kpc could have different causes: (1) local dynamics of the spiral structure, (2) resonance of the spiral structure, or (3) resonance of the bar potential. 

\citet{quillen11} argued that, within a density wave, the orientations of the orbits of the stars on the spiral arm are correlated with the mean radii of the orbits and result in an arc-like structure in the $U-V$ velocity distribution. Since what we see is only the marginalized distribution over $V$, an arc would be difficult to show. Even if the arc is elongated along $V$ and thus show a sharp peak in $U$, it is hard to explain the $\sim$30\,\kms\ gap between the two peaks. Moreover, the bifurcation is at a larger radius than the location of the Perseus arm. Therefore, it seems that the local orbital distribution in the spiral arm cannot be the main reason of the bifurcation. 

If 10--11\,kpc corresponds to the corotation radius of the spiral structure then its pattern speed is 20--22\,km s$^{-1}$\,kpc$^{-1}$, in agreement with \citet{gerhard11}. However, \citet{antoja11} show that spiral structure alone results typically in too little substructure in the velocity distribution.

The bifurcation can be the result of a resonance with the Galactic bar. 
The pattern speed of the Galactic bar has been reported in the range from 30 to 65\,km s$^{-1}$\,kpc$^{-1}$ depending on different data and methods \citep{gerhard11}. If we adopt 40\,km s$^{-1}$\,kpc$^{-1}$ and a flat rotation curve of $v_c=220$\,km s$^{-1}$ \citep{Rod08}, the OLR radius is right around 10--11\,kpc.
 It is then expected that stars group into inward and outward streams under the OLR radius, naturally explaining the bifurcation in $RV$.
Meanwhile, the CR radius corresponding to this pattern speed is located around 5-6\,kpc, which can explain the Hercules stream in the Solar neighborhood. 
As even the bar pattern speed itself is not yet well determined, further investigation in both observations and theory are needed to verify the resonant origin of the bifurcation. 

\section{Summary}
\label{s:summary}

In this letter, we use 698 red clump stars in the mid-plane of the Milky Way to derive the radial profile of extinction and radial velocity dispersion, as well as the velocity-distance distribution in the anti-center direction. We find:

i) Dustlanes in front of the Perseus and Outer arms, consistent with external spiral galaxies;

ii) A Perseus arm-interarm contrast between 1.3 and 1.5, in line with previous studies;

iii) A strong bifurcation in the radial velocity distribution around 10--11\,kpc, likely related to the outer Lindblad resonance of the Galactic bar, although additional observations are needed to constrain a possible spiral corotation resonance.

\acknowledgments
The authors thank Shude Mao for his very helpful comments. This research uses data obtained through the Telescope Access Program (TAP), which is funded by the National Astronomical Observatories and the Special Fund for Astronomy from the Ministry of Finance. X. Xue also acknowledges the Young Researcher Grant of National Astronomical Observatories, Chinese Academy of Sciences. This work was supported by Chinese National Natural Science Foundation
(NSFC) through grant No.11243003, Y111181001, 11150110135 and Sonderforschungsbereich SFB 881 ``The Milky Way System" (subproject A7) of the German Research Foundation (DFG).


\end{document}